# On the role of extrinsic noise in microRNA-mediated bimodal gene expression

Marco Del Giudice,[1,2,*] Stefano Bo,[3] Silvia Grigolon,[4,†] and Carla Bosia [*,†,1,2]

[1]*Department of Applied Science and Technology, Politecnico di Torino,*

*Corso Duca degli Abruzzi 24, I-10129 Torino, Italy*

[2]*Human Genetics Foundation, via Nizza 52, I-10126 Torino, Italy*

[3]*Nordita, Royal Institute of Technology and Stockholm University,*

*Roslagstullsbacken 23, SE-10691 Stockholm, Sweden*

[4]*The Francis Crick Institute, 1 Midland Road, London NW1 1AT, UK*

Several studies highlighted the relevance of extrinsic noise in shaping cell decision making and differentiation in molecular networks. Experimental evidences of phenotypic differentiation are given by the presence of bimodal distributions of gene expression levels, where the modes of the distribution often correspond to different physiological states of the system. We theoretically address the presence of bimodal phenotypes in the context of microRNA (miRNA)-mediated regulation. MiRNAs are small noncoding RNA molecules that downregulate the expression of their target mRNAs. The nature of this interaction is titrative and induces a threshold effect: below a given target transcription rate no mRNAs are free and available for translation. We investigate the effect of extrinsic noise on the system by introducing a fluctuating miRNA-transcription rate. We find that the presence of extrinsic noise favours the presence of bimodal target distributions which can be observed for a wider range of parameters compared to the case with intrinsic noise only and for lower miRNA-target interaction strength. Our results suggest that combining threshold-inducing interactions with extrinsic noise provides a simple and robust mechanism for obtaining bimodal populations not requiring fine tuning. We furthermore characterise the protein distributions dependence on protein half-life.

* To whom the correspondence should be addressed: marco.delgiudice@polito.it, carla.bosia@polito.it.

† These authors contributed equally.

# I. INTRODUCTION

Environmental fluctuations can be source of noise in molecular networks, usually referred to as extrinsic noise [1, 2]. Extrinsic noise, together with intrinsic fluctuations due to the probabilistic nature of chemical reactions, shapes gene expression and thus cell decision making and differentiation. Cells fate is normally triggered by non-trivial patterns of chemicals, such as morphogens, which are differentially expressed throughout the tissues. Cell phenotypic variability can be therefore measured by quantifying the underlying distribution of the triggering chemicals or of the downstream proteins in the same pathway. Bimodal distributions turn out to be a common outcome of gene-expression data, ranging from cancer to immune cells [3, 4]. The two peaks of the distribution are usually associated with different physiological states of the system, may they refer to different organ primordia or different disease states or cancer subtypes [5–8].

In the past, some studies highlighted the possibility that microRNAs (miRNAs), similarly to their bacterial counterpart [9], may induce bimodality in the expression of their targets thanks to their specific titrative interactions [10–12].

MiRNAs are small molecules of non-coding RNA found in eukaryotes to act as post-transcriptional regulators. Although they were found in several different eukaryotic kingdoms, their role is known to be vital in multicellular organisms. They perform this function by recognising mRNA targets through Watson-Crick base pairing. Once bound to the target, they can both enhance the instability of that mRNA by degrading it and decrease its translation by keeping it bound. Interestingly, different levels of interaction miRNA-target can be achieved by different numbers of miRNAs [13–15]. Theoretical predictions together with in vitro single-cell experiments suggested that bimodality in the expression level of miRNA targets can be achieved with a high miRNA-target interaction strength [11, 12]. In terms of genetic sequences, this would imply a high specificity between target and miRNA and therefore a high number of complementary binding sites (bs) per target.

As long as a miRNA molecule is bound to the target, such target cannot be translated. It is then possible to define a threshold for the mRNA transcription rate such that below the threshold all the mRNA target molecules are bound to miRNAs and above the threshold there are molecules of mRNA free for translation [11, 16, 17]. Close to the threshold the number of both free miRNAs and targets is small, their fluctuations are highly coupled thanks to the non-linear interaction between the two and a small fluctuation in their amount of molecules may lead the system from the "bound" to the "unbound" state [11, 12, 17]. If the interaction strength between miRNA and target is high, then the transition from the bound to the unbound state is sharp. Close to the threshold, simply because of intrinsic fluctuations in the amount of both miRNA and target, part of the targets will be bound to the miRNA and part unbound for the same transcription rate. Picturing this in terms of target distribution would lead to a bimodal distribution whose two modes are associated to the bound and unbound state. It is worth to underline that this kind of bimodality is due to the presence of noise and not to peculiar molecular mechanisms introducing multiple deterministic stable states in the system.

MiRNAs are predicted to regulate more than 60% of our genome through a combinatorial action: every single miRNA can regulate several targets and one target can be regulated by different miRNAs [15, 18]. The variety of targets they regulate is so wide and important for different signalling pathways or developmental stages [19, 20] that the alteration of their expression levels may contribute to diseases such as tumour development and metastatisation [21–24]. It is nowadays also well established that multiple cell-cycle regulators are controlled by miRNAs, whose regulation could be in turn cell-cycle dependent [25–28]. The expression level of miRNAs may thus change with the cell-cycle progression, and there are indeed miRNAs differentially expressed according to the particular phase of the cell cycle [29]. As a consequence, in a population of cells heterogeneous with respect to the cell cycle, such as non-quiescent cancer cells, the amount of miRNAs can strongly fluctuate from cell to cell. This introduces an extra source of noise in the system besides the intrinsic stochasticity of chemical reactions involving gene transcription, translation and regulation.

Here we study, both analytically and with simulations, how this extrinsic source of noise influences miRNA-mediated regulation. Although extrinsic noise affects also the expression of the other species in the network, we are here interested in specifically understanding the effects of a noisy source on miRNAs. We show how a distribution of miRNA transcription rates reshapes the threshold between miRNA and target and defines a wide region of bimod-ality. Such a bimodal distribution can be seen at a "population level", being the amount of miRNA heterogenous throughout the different cells. This outcome is deeply different than previous results where differential phenotypic expression is induced by the strong coupling between miRNA and its target at the "single-cell level". Interestingly, we also show that, if the miRNA target is protein coding, the protein half-life can alter the protein distribution. With respect to the shape of the mRNA distribution, an increased protein half-life leads to a narrowing of the protein distribution around its mean. This may promote or suppress bimodality, suggesting that bimodal distributions at the level of mRNA may still correspond to a specific single phenotype at the protein level. Conversely, repressed heavy tailed mRNA distributions may give rise to bimodal protein distributions.

Finally, given the existence of multiple targets competing for one type of miRNA, we ask whether these properties can be maintained in a more complex circuit made of two competing endogenous RNAs (ceRNAs) and one miRNA [30]. The different target genes act indeed as sponges for the miRNA molecules and may sequester them from the environment. As a result, the overexpression or underexpression of one of the targets can lead respectively to an increase or decrease in the level of expression of the other competitors. The intensity of such cross regulation depends on the distance from the threshold of quasi equimolarity between miRNAs and targets [11, 17]. This suggests that, if one target has a bimodal distribution, such bimodality may be influenced by the expression levels of the other miRNA competitors.

## II. MATERIALS AND METHODS

### A. Stochastic model for miRNA-target interaction with extrinsic noise

Models of microRNA-mediated circuits have been the subject of several studies over the past years [16, 34, 35]. Hereby, we will refer to one of the simplest ways of accounting for microRNA-driven inhibition, depicted in Figure 1A. The molecular species involved in this circuit are miRNAs (S), target mRNAs (R) and proteins (P), result of the translation of the target mRNA.

In the following, we shall assume miRNAs and mRNAs to be transcribed from independent genes. For simplicity we neglect all the intermediate reactions leading to the synthesis of mRNAs, therefore assuming they are produced at constant rate $k_R$. For the miRNA we consider it to be transcribed with a constant rate $k_S$ which we let fluctuate between different cells to probe the effects of extrinsic noise on the system. MiRNAs and mRNAs can also be degraded by the action of specialised enzymes. Hereby, we assume these reactions to be governed by mass-action law with rates $g_S$ and $g_R$. The associated molecular reactions read:

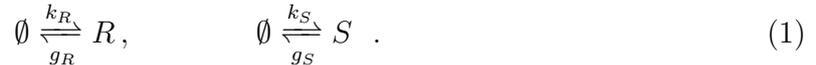

$$\emptyset \underset{g_R}{\overset{k_R}{\rightleftharpoons}} R\,, \qquad \emptyset \underset{g_S}{\overset{k_S}{\rightleftharpoons}} S \quad . \tag{1}$$

MiRNAs act as post-transcriptional regulatory elements, by binding the target mRNAs in a complex that can be subsequently degraded. Such interaction between miRNAs and mRNAs is quantified by the effective parameter $g$, which takes into account the strength of the coupling miRNA-target: from a biochemical point of view, it depends on the number of miRNA binding sites dedicated to a specific target [16]. The formation of miRNA-mRNA complex reads:

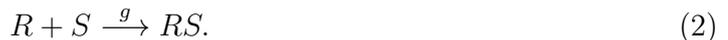

$$R + S \overset{g}{\longrightarrow} RS. \tag{2}$$

While the mRNAs are always degraded due to the titrative interaction, the miRNAs can be recycled with probability $1 - \alpha$ in the following way:

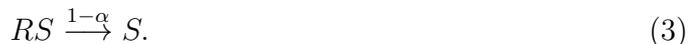

$$RS \overset{1-\alpha}{\longrightarrow} S. \tag{3}$$

Whenever the mRNAs are not bound to miRNAs, they can be translated into proteins with translation rate $k_P$ and, as assumed for the other molecular species, proteins can be as well degraded with rate $g_P$, i.e.:

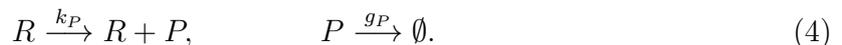

$$R \overset{k_P}{\longrightarrow} R + P, \qquad P \overset{g_P}{\longrightarrow} \emptyset. \tag{4}$$

From now on, we define as "intrinsic noise" the fluctuations due to the stochasticity of the chemical reactions with constant rates (Figure 1A) and as "extrinsic noise" those due to the fluctuating miRNA transcription rate (see Figure 1C).

The system can be described by the probability distribution $P(n_S, n_R, n_P, t|\mathcal{K})$ of observing $n_S$ molecules of miRNA, $n_R$ molecules of mRNA and $n_P$ proteins at time $t$ given a set of parameters $\mathcal{K} = \{k_R, k_S, k_P, g_R, g_S, g_P, g, \alpha\}$. This probability distribution follows the same master equation presented in [11] that can be either solved numerically or at the steady-state with some approximations. If the parameters fluctuate, in order to obtain the full distribution at the steady state $P_{ss}(n_S, n_R, n_P)$ one has to take into account such fluctuations. This can be achieved by using the law of total probability [32], which states that $P(n_S, n_R, n_P) = \int P(\mathcal{K})P(n_S, n_R, n_P|\mathcal{K})d\mathcal{K}$. As our aim is to test the effects of a fluctuating miRNA transcription rate, we shall assume this to be the only parameter drawn from a probability distribution, specifically a Gaussian centred around $\langle k_S \rangle$ with variance $\sigma_{k_S}$ and defined only for positive values of $k_S$.

To obtain the steady-state distribution $P(n_S, n_R, n_P|k_S)$ conditioned on a specific miRNA transcription rate we could chose different approximation methods. Pivotal examples are the Van Kampen [31] and the gaussian approximations [11]. In the following we focused on the first one, leaving to the Supplementary Information (SI) a comparison between the two methods. We therefore performed a system-size expansion, thus assuming the system distribution at fixed parameters to be gaussian.

The marginal distribution $P(n_S, n_R, n_P)$ is then found by using the law of total probability, i.e., by performing a weighted average on all possible values of $k_S$.

Finally, we applied the same approach when considering two targets interacting with the same miRNA (Figure 5A). In this case the conditional distribution is $P(n_S, n_{R1}, n_{R2}, n_{P1}, n_{P2}|k_S)$ from which one can obtain the full one by integrating over the values of the miRNA transcription rate.

## Analytical approach

The equations governing the dynamics of the system considered in Figure 1 are given by:

$$\begin{aligned}
\frac{dR}{dt} &= k_R - g_R R - g R S \\
\frac{dS}{dt} &= k_S - g_S S - g \alpha R S \\
\frac{dP}{dt} &= k_P R - g_P P \quad,
\end{aligned} \tag{5}$$

where $R, S$ and $P$ are the concentrations of the three species involved in the circuit expressed in nanomolars and the parameters are the same as in Figure 1.

Because of the inherent stochastic nature of molecular reactions, intrinsic noise should be taken into account by defining the probability distribution of observing $\underline{n} = (n_R, n_S, n_P)$ molecules at time $t$, namely $P(\underline{n}, t)$. The number of molecules of the species $X$, $n_X$, relates to the concentrations, $\rho_X$, as $n_X = V_{cell}\rho_X$. The dynamics of this system can be rewritten

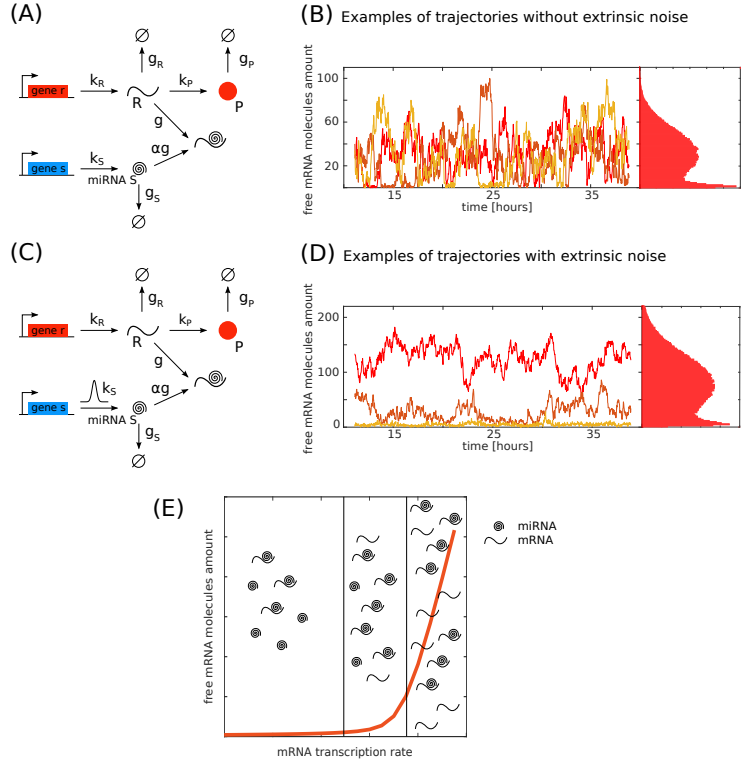

Figure 1: **Model and steady-state trajectories.** The reference circuits (without and with extrinsic noise on the miRNA transcription rate) are represented in (A) and (C) respectively with the rates considered in the model. $k_R$ and $k_S$ are the target mRNA and miRNA transcription rates, $g_R$ and $g_S$ are respectively the mRNA and miRNA degradation rates. $k_P$ is the protein translation rate and $g_P$ is its degradation rate. $g$ is the miRNA-target interaction strength and $\alpha$ is the fraction of miRNAs that are not recycled after binding to the mRNA. In panels (B) and (D) there are three different trajectories for the mRNA, corresponding to the model on the left. For both panels, the steady-state distributions of the number of free mRNA molecules are bimodal. In (B) the parameters are $k_S = 1.2 \times 10^{-3}$ nM min$^{-1}$, $k_R = 2.7 \times 10^{-3}$ nM min$^{-1}$, $g_S = 1.2 \times 10^{-2}$ min$^{-1}$, $g_R = 2.4 \times 10^{-2}$ min$^{-1}$, $g = 1.5 \times 10^{3}$ nM$^{-1}$ min$^{-1}$, $k_P = 6.0$ min$^{-1}$, $g_P = 1.2 \times 10^{-2}$ min$^{-1}$ and $\alpha = 0.5$. In (D) the parameters are $k_R = 3.1 \times 10^{-3}$ nM min$^{-1}$, $g_S = 1.2 \times 10^{-2}$ min$^{-1}$, $g_R = 2.4 \times 10^{-2}$ min$^{-1}$, $g = 1.2 \times 10^{2}$ nM$^{-1}$ min$^{-1}$, $k_P = 6.0$ min$^{-1}$, $g_P = 1.2 \times 10^{-2}$ min$^{-1}$ and $\alpha = 0.5$. $k_S$ are picked from a gaussian distribution with mean $\overline{k}_S = 1.2 \times 10^{-3}$ nM min$^{-1}$ and standard deviation $\sigma = 2.4 \times 10^{-4}$ nM min$^{-1}$. (E) Cartoon of the free mRNA threshold behaviour as a function of its transcription rate. Below the threshold the amount of free miRNA is greater then the amount of free mRNA, in proximity to the threshold their amount is nearly the same, above the threshold the mRNA free amount exceeds the miRNA.

in terms of the master equation, that reads:

$$\frac{dP(\underline{n}, t)}{dt} = k_R \Big[ P(n_R - 1, t) - P(\underline{n}, t) \Big] +$$

$$\frac{g_R}{V_{cell}} \Big[ (n_R + 1) P(n_R + 1, t) - n_R P(\underline{n}, t) \Big] +$$

$$k_S \Big[ P(n_S - 1, t) - P(\underline{n}, t) \Big] +$$

$$\frac{g_S}{V_{cell}} \Big[ (n_S + 1) P(n_S + 1, t) - n_S P(\underline{n}, t) \Big] +$$

$$\frac{k_P n_R}{V_{cell}} \Big[ P(n_P - 1, t) - P(\underline{n}, t) \Big] +$$

$$\frac{g_P}{V_{cell}} \Big[ (n_P + 1) P(n_P + 1, t) - n_P P(\underline{n}, t) \Big] +$$

$$\frac{g \alpha}{V_{cell}^2} \Big[ (n_S + 1)(n_R + 1) P(n_R + 1, n_S + 1, t) - n_S n_R P(\underline{n}, t) \Big] +$$

$$\frac{g(1 - \alpha) n_S}{V_{cell}^2} \Big[ (n_R + 1) P(n_R + 1, t) - n_R P(\underline{n}, t) \Big]. \tag{6}$$

Solving the master equation or even only determining the first and second moments of $P(\underline{n}, t)$ might be a difficult task due to the non-linear terms appearing from the miRNA-mRNA interactions. It is common practice to refer to approximations, amongst all the van Kampen's system-size expansion [31].

The van Kampen's system-size expansion relies on the assumption that the number of molecules of any species $X$ can be split into two contributions, the former coming from the deterministic system and the latter from the intrinsic noise in the system, namely:

$$n_X = V_{cell} \rho_X + V_{cell}^{1/2} \xi_X, \tag{7}$$

where $\xi_X$ is a gaussian-distributed variable with zero average. The cell volume, $V_{cell}$, represents the system size and it is assumed to be sufficiently large. Rewriting the master equation by using Eq. 7, at the leading order in $V_{cell}$ one finds the probability distribution to be a Dirac's delta in terms of the number of molecules, therefore recovering the deterministic equations Eqs. **5**. The next-to-leading order in the expansion instead leads to a linear Fokker-Planck like equation in terms of the noisy variables $\{\xi_X\}$, the probability distribution of the number of molecules then being gaussian centred around the deterministic concentrations $\phi_X$ and with finite variance. By using this equation, one can compute the variances and the cross-correlations of the noisy variables, namely $\langle \xi_X^2 \rangle$ and $\langle \xi_X \xi_Y \rangle$ that can be related to the molecules' variances and cross-correlations.

In the specific case of the circuit we analysed, the variances and cross-correlations are steady-state solutions of the following system:

$$\langle \dot{\xi_R^2} \rangle = k_r - 2g_r\langle \xi_R^2 \rangle + g_r R - 2\alpha gR\langle \xi_S \xi_R \rangle - 2\alpha gS\langle \xi_R^2 \rangle +$$

$$\alpha gSR - 2(1-\alpha)gS\langle \xi_R^2 \rangle + (1-\alpha)gRS - 2(1-\alpha)gR\langle \xi_S \xi_R \rangle$$

$$\langle \dot{\xi_S^2} \rangle = k_s - 2g_s\langle \xi_S^2 \rangle + g_s S - 2\alpha gR\langle \xi_S^2 \rangle - 2\alpha gS\langle \xi_S \xi_R \rangle + \alpha gRS$$

$$\langle \dot{\xi_P^2} \rangle = k_p R - 2g_p\langle \xi_P^2 \rangle + g_p P + 2k_p\langle \xi_P \xi_R \rangle$$

$$\langle \dot{\xi_R \xi_S} \rangle = -g_r\langle \xi_S \xi_R \rangle - g_s\langle \xi_S \xi_R \rangle - \alpha gR\langle \xi_S^2 \rangle - \alpha gS\langle \xi_S \xi_R \rangle -$$

$$\alpha gR\langle \xi_S \xi_R \rangle - \alpha gS\langle \xi_R^2 \rangle + RS\alpha g - (1-\alpha)gS\langle \xi_S \xi_R \rangle -$$

$$(1-\alpha)gR\langle \xi_S^2 \rangle$$

$$\langle \dot{\xi_R \xi_P} \rangle = -g_p\langle \xi_P \xi_R \rangle + k_P\langle \xi_R^2 \rangle - \alpha gR\langle \xi_P \xi_S \rangle - \alpha gS\langle \xi_P \xi_R \rangle -$$

$$(1-\alpha)gS\langle \xi_P \xi_R \rangle - (1-\alpha)gR\langle \xi_P \xi_S \rangle - g_r\langle \xi_P \xi_R \rangle$$

$$\langle \dot{\xi_S \xi_P} \rangle = -g_p\langle \xi_P \xi_S \rangle - g_s\langle \xi_P \xi_S \rangle + k_P\langle \xi_S \xi_R \rangle - \alpha gR\langle \xi_P \xi_S \rangle -$$

$$\alpha gS\langle \xi_P \xi_R \rangle \quad . \tag{8}$$

Using the expressions obtained from the previous system, specifying all the parameters, one can compute the probability distribution of the system, $P(\underline{n}, t)$.

To investigate the role of environmental fluctuations (i.e. extrinsic noise) in our system of interest, we consider a fluctuating rate of miRNA production $k_S$. For the sake of simplicity, we assume this parameter to be drawn from a gaussian distribution, $P(k_S) = \frac{1}{\sqrt{2\pi\sigma_{k_S}^2}} e^{-\frac{(k_S - \langle k_S \rangle)}{2\sigma_{k_S}^2}}$, defined for positive values of $k_S$, where $\langle k_S \rangle$ and $\sigma_{k_S}$ are respectively the average and variance of $k_S$.

Due to the further stochasticity introduced by the extrinsic noise, the master equation previously derived does not hold anymore. Fluctuations on the parameter $k_S$ do not allow to rewrite a similar equation for this system in a simple way. However, as also shown in [32], the probability distribution of the entire system, $P(\underline{n})$ can be rewritten in terms of conditional probabilities by using the law of total probability in the following way:

$$P(\underline{n}) = \int P(k_S)P(\underline{n}|k_S)dk_S, \tag{9}$$

where $P(k_S)$ is the gaussian distribution in $k_S$ and $P(\underline{n}|k_S)$ is the conditioned probability of observing a certain configuration of the system, $\underline{n}$, given a specific value of $k_S$. This probability distribution is solution of the master equation Eq. 6 for any given $k_S$. One can therefore apply again the van Kampen's expansion on the master equation of $P(\underline{n}|k_S)$ and obtain all the moments of this distribution (that will be all functions of the fluctuating parameter $k_S$). The full solution can be obtained by averaging the result over all the values of $k_S$.

**Molecular simulations of microRNA mediated circuits via the Gillespie's algorithm**

Stochastic simulations have been performed via implementation of the Gillespie's direct algorithm [33]. All the results presented in this paper are obtained at the steady state.

## III. RESULTS

### A. Single cell versus population-induced bimodality

As discussed in the Introduction, the understanding of bimodal distributions is usually related to cell fate determination and differentiation. These mechanisms are at the basis of organism development and mis-development. It is therefore important to address the question about which might be the underlying molecular mechanisms allowing cell diversity and variability. Given the strong involvement of miRNAs in developmental stages, we focus here on the miRNA network represented in Figure 1A, at the single-cell level. Previous works [16, 30] showed that the binding and unbinding reactions between miRNA and target give rise to non-trivial threshold effects in condition of quasi equimolarity between miRNA and target, see Figure 1E, where the threshold is defined in terms of miRNA and mRNA transcription rates as $k_S^* = \alpha k_R^*$ [16]. If the miRNA is transcribed at a rate above the threshold value, $k_S > k_S^*$, the system is enriched in microRNA, which tends to bind most of the present mRNA therefore preventing its translation. In this regard, the system can be seen as below the threshold with respect to the target and we shall refer to it as in the "repressed state". Above the threshold, the mean amount of free target increases linearly with its transcription rate. The scenario with free mRNA molecules will be denoted as the "unrepressed state" of the system. Such threshold effect displaying a rather sharp transition between the repressed and the unrepressed state gets more marked as the interaction strength between miRNA and targets increases. Close to the threshold value of the target transcription rate, due to the probabilistic nature of chemical reactions, the system will stochastically switch between the repressed and unrepressed state. This stochastic switching is enough to give rise to bimodal target distributions which appear for a narrow range of the target transcription rate $k_R$ [11] (Figure 1B). Such bimodality characterises the single cell where the miRNA network is defined: every single cell can jump from the repressed to the expressed target state if the coupling constant with the miRNA is high enough.

On the contrary, in presence of extrinsic noise, the miRNA transcription rate is not the same for every cell (Figure 1C,D). Hereby, we model the extrinsic noise through a gaussian-distributed miRNA transcription rate. To understand intuitively the consequences of this kind of extrinsic noise, let us consider the case of a miRNA transcription-rate distribution with fixed average $\langle k_S \rangle$. When the mRNA transcription rate ($k_R$) is very low and the average miRNA transcription rate is much larger than the threshold value ($\langle k_S \rangle \gg \alpha k_R$), most of the drawn transcription rates $k_S$ will be larger than the threshold value. This would place the network in the parameter range where the targets are all bound to the miRNAs (see Figure 2). For larger $k_R$, approaching the threshold, values of $k_S$ extracted from the left-tail will correspond to the case with some unbound targets. Below the threshold, as $\langle k_S \rangle < \alpha k_R$, the majority of the drawn $k_S$ will belong to the unrepressed state with the right tail of the distribution sampling from the all-bound region (Figure 2A,B). This scheme however is the same as before at a population level. The presence of rates above and below the threshold across the population can give rise therefore to a bimodal distribution in the number of free targets (Figure 2A,C). In particular, the higher the extrinsic noise, the larger the range of target transcription rates for which bimodality is present and the greater the separation between the two phenotypes (bound and unbound targets), see Figure 3A. This implies that, in contrast to the case without extrinsic noise, it is no longer necessary to fine tune the transcription rates to obtain a bimodal distribution. Even for high values of $k_R$, the fraction of randomly picked $k_S$ that brings the system in the bound state is not negligible and forms a visible peak in the distribution (Figure 2A-C). The bimodal distribution is in this case given by the superposition of unimodal distributions obtained for different $k_S$ and weighted by the probability $P(k_S)$ (see Figure S2 in SI). Focusing on one particular value of the variance $\sigma_{k_S}$ and varying the target transcription rate, $k_R$, we monitored the appearance of bimodal distributions. We ran Gillespie's simulations from which we sampled the number of targets for the histograms shown in Figure 2. By using system-size expansion and the law of total probability, we obtained the analytical targets distributions, shown in Figure

2. The analytical approximation captures the behaviour of the system for the mean, the Coefficient of Variation and the probability distribution, as testified by the agreement with the simulations (see Figure 2A,C,D and S3 in SI).

## B. A noisy environment can compensate for low miRNA-target interaction to obtain bimodal distribution

Let us focus in more detail on the effects of the extrinsic noise on the bimodal properties of miRNA networks. To dissect the properties of bimodal distributions, we first ran Gillespie's algorithm simulating the network in Figure 1C for different target transcription rates, $k_R$, and different variances of the Gaussian noise on the miRNA transcription rates $\sigma_{k_S}$. Monitoring the appearance of bimodal distribution, one can build up a phase diagram like the one represented in Figure 3A. In absence of extrinsic noise but presence of intrinsic noise, bimodal distributions appear only for high coupling between miRNA and target and this region gets wider upon increasing the coupling constant $g$. Therefore the interaction strength between miRNA and target $g$ affects the range of values of $k_R$ in which the bimodality is present.

Since in this case the bimodality is a single-cell effect, only those cells having the target highly interacting with the miRNA have chance to experience the repressed and unrepressed state when $k_R$ is close to its threshold value.

Adding some extrinsic noise relaxes the constraint on the interaction strength. Bimodality becomes rather a population effect, with some cells being locked in the repressed state (by having large miRNA transcription rates $k_S$) and some others (with smaller $k_S$) displaying free targets. Figure 3B shows how it is possible to have similar bimodal distributions either increasing the miRNA-target interaction strength (blue histogram) or increasing the extrinsic noise (orange histogram) with respect to a reference case with pure intrinsic noise and low miRNA-target interaction (black line). The extrinsic noise and the miRNA-target interaction strength act at a similar level with respect to bimodality, where a higher extrinsic noise can compensate for a low interaction strength (small number of miRNA binding sites on the target) in order to obtain two differentially expressed phenotypes.

**(A)   miRNA production rate distributions**

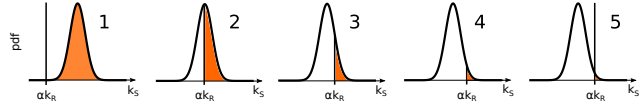

**(B)   mRNA mean**

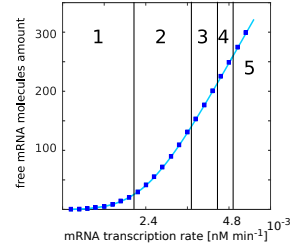

**(C) free mRNA molecules distributions**

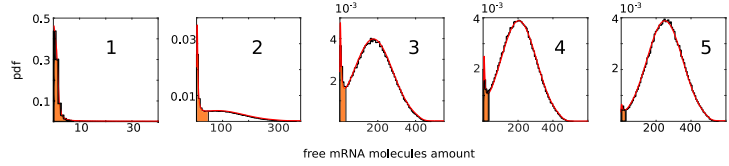

**(D) protein molecules distributions**

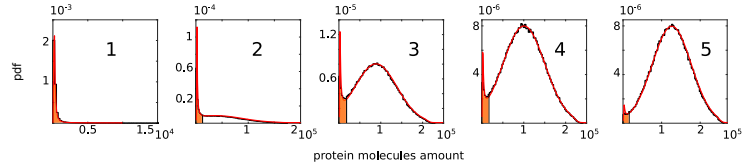

Figure 2: **Emergence of bimodality in presence of extrinsic noise.** (A) Qualitative representation of the miRNA production rate distribution. The black vertical line indicates the value of the miRNA transcription rate $k_S = \alpha k_R$ for different values of the target transcription rate $k_R$. The distributions represent the different regions labelled from 1 to 5 shown in (B). The region of the distribution contributing to the repressed state is coloured in orange. (B) Free mRNA molecules amount as a function of the target transcription rate $k_R$. Solid lines are analytical predictions while blue squares correspond to numerical simulations. (C) Free mRNA molecules distributions corresponding to the regions in (B). Solid black lines correspond to numerical simulations while solid red lines are analytical predictions. The repressed region is coloured in orange. (D) Protein molecules distributions corresponding to the mRNA distributions in (C). Solid black lines correspond to numerical simulations while solid red lines are analytical predictions. The repressed region is coloured in orange. In (B) the parameters are $g_S = 1.2 \times 10^{-2}$ min$^{-1}$, $g_R = 2.4 \times 10^{-2}$ min$^{-1}$, $g = 1.2 \times 10^3$ nM$^{-1}$ min$^{-1}$, $k_P = 6.0$ min$^{-1}$, $g_P = 1.2 \times 10^{-2}$ min$^{-1}$, $\alpha = 0.5$. $k_S$ are picked from a gaussian distribution with mean $\overline{k}_S = 1.2 \times 10^{-3}$ nM min$^{-1}$ and standard deviation $\sigma = 4.8 \times 10^{-4}$ nM min$^{-1}$. $k_R$ ranges from $2.4 \times 10^{-4}$ nM min$^{-1}$ to $5.2 \times 10^{-3}$ nM min$^{-1}$. In (C) and (D) the parameters are the same as in (B) except for $k_R$ that is fixed for each distribution and from region 1 to 5 takes the values $k_R = 1.2 \times 10^{-3}$ nM min$^{-1}$, $3.1 \times 10^{-3}$ nM min$^{-1}$, $4.0 \times 10^{-3}$ nM min$^{-1}$, $4.3 \times 10^{-3}$ nM min$^{-1}$, $4.8 \times 10^{-3}$ nM min$^{-1}$.

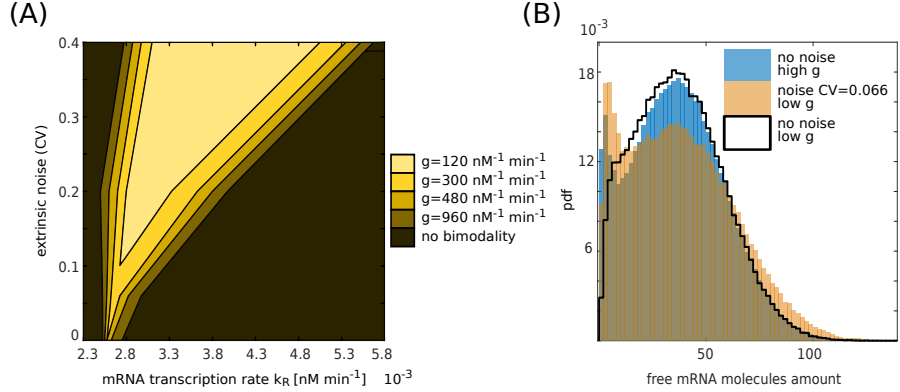

Figure 3: **Bimodality as a function of the parameters.** (A) Phase diagram for bimodality in the free mRNA molecules distribution. On the $x$ axes there is the target transcription rate $k_R$, on the $y$ axes the extrinsic noise on the miRNA transcription rate. The color map indicates the presence of bimodality for different values of the miRNA-target interaction strength $g$. The width of the bimodality range increases as the interaction strength or the extrinsic noise are increased. The phase diagram is the result of a series of simulations in which the presence of bimodality was checked manually. The following parameters are equal for all the simulations: $g_S = 1.2 \times 10^{-2}$ min$^{-1}$, $g_R = 2.4 \times 10^{-2}$ min$^{-1}$, $k_P = 6.0$ min$^{-1}$, $g_P = 1.2 \times 10^{-2}$ min$^{-1}$, $\alpha = 0.5$. Target mRNA transcription rate is one of the control parameters and ranges from $k_R = 2.4 \times 10^{-3}$ nM min$^{-1}$ to $k_R = 5.7 \times 10^{-3}$ nM min$^{-1}$. miRNA-mRNA association rate is one of the control parameters and takes the following values: $g = 1.2 \times 10^2$ nM$^{-1}$ min$^{-1}$, $3.0 \times 10^2$ nM$^{-1}$ min$^{-1}$, $4.8 \times 10^2$ nM$^{-1}$ min$^{-1}$, $9.7 \times 10^2$ nM$^{-1}$ min$^{-1}$. Extrinsic noise is tuned by varying the standard deviation of the distribution with mean $\overline{k}_S = 1.2 \times 10^{-3}$ nM min$^{-1}$ from which miRNA transcription rates are picked, the standard deviation takes the following values: $\sigma = 0$ nM min$^{-1}$ (no extrinsic noise), $7.1 \times 10^{-5}$ nM min$^{-1}$, $2.4 \times 10^{-4}$ nM min$^{-1}$, $4.8 \times 10^{-4}$ nM min$^{-1}$. To define the origin of the bimodality region for the case with $g = 1.2 \times 10^2$ nM$^{-1}$ min$^{-1}$ the value $\sigma = 1.2 \times 10^{-4}$ nM min$^{-1}$ was also used. (B) Free mRNA distribution in case of pure intrinsic noise and small ($g = 3.8 \times 10^2$ nM$^{-1}$ min$^{-1}$) miRNA-target interaction strength (black line), pure intrinsic noise and high ($g = 1.1 \times 10^3$ nM$^{-1}$ min$^{-1}$) miRNA-target interaction (blue histogram) and extrinsic noise ($\sigma = 7.9 \times 10^{-5}$ nM min$^{-1}$) and small ($g_1 = 3.8 \times 10^2$ nM$^{-1}$ min$^{-1}$) miRNA-target interaction strength (orange histogram). The other parameters are as follows: $k_S = 1.2 \times 10^{-3}$ nM min$^{-1}$, $g_S = 1.2 \times 10^{-2}$ min$^{-1}$, $k_R = 2.7 \times 10^{-3}$ nM min$^{-1}$, $g_R = 2.4 \times 10^{-2}$ min$^{-1}$, $k_P = 6.0$ min$^{-1}$, $g_P = 1.2 \times 10^{-2}$ min$^{-1}$, $\alpha = 0.5$. The plot shows how extrinsic noise can compensate for small miRNA-target interaction strength in order to obtain bimodal distributions.

## Interplay between different targets increases the stability of bimodal phenotypes

The study so far led us to picture the possible importance of extrinsic noise in cell phenotypic variability. Given the existence of multiple miRNA-targets networks, we now investigate how the results for the one-target case, extend to the multiple-target one. Let us consider a minimal model with two targets, $R_1$ and $R_2$, competing for the same miRNA, $S$ (Figure 4A). We start by investigating the effect of an increase in the expression of target $R_2$ on target $R_1$ with and without extrinsic noise. Upon increasing the transcription rate $k_{R_2}$ of target $R_2$, the threshold of the target $R_1$ shifts towards lower expression levels: the miRNAs are indeed sponged away by $R_2$ and a lower amount of $R_1$ is needed to overcome the threshold.

If $R_1$ has a high interaction strength $g_1$ with the miRNA, then the range of bimodality shifts towards lower expression levels as well. The width of the range of $R_1$ bimodality is determined by the interaction strength $g_2$ of the target $R_2$ with the miRNA. If $g_2 >> g_1$, then the miRNAs are sponged away by the second target with such a high frequency that the net effect is a reduction in the amount of miRNAs available to target $R_1$. This entails a shift not only of the $k_{R_1}$ threshold value but also of the range of bimodality. If $g_2 < g_1$, the second target $R_2$ interacts with low frequency with the miRNA, $R_1$ is slightly derepressed and the net effect on its bimodal distribution is a shrink of the range of transcription rates for which it is present. The emerging picture is that, for a given transcription rate $k_{R_1}$, it is possible to tune the distribution of target $R_1$ from monomodal to bimodal and from unrepressed to repressed and viceversa via the expression of target $R_2$.

The presence of extrinsic noise makes such cross regulation possible also for cases with lower miRNA-target interaction on both targets. In Figure 4B we show the bimodality phase diagram for $R_1$ at a fixed interaction strength $g_1$ between miRNA and target $R_1$, and for a fixed level of extrinsic noise (see Figure S4 in SI for a different noise level). The interaction strength is such that in case of pure intrinsic noise $R_1$ does not show bimodal distribution. As explanatory example, Figure 4C shows that the peaks of $R_1$ distribution can be tuned towards the repressed or the unrepressed case decreasing or increasing the expression of a second target $R_2$. Here, the two targets $R_1$ and $R_2$ are both coupled through the noisy miRNA with small interaction strengths. Also in this case, the same analytical approach as before gives good agreement between theory and simulations.

These observations suggest that even if the miRNA repression is low and diluted over a network of multiple targets, the noisy environment allows cross-regulation between ceRNAs at the population level (see Figure 5D, with the intersection between the bimodality phase diagrams for both targets for a fixed level of noise and miRNA interaction strengths).

**Protein stability and bimodal phenotypes**

Given the relevance of the final product of gene expression, it is important to consider what is the effect of extrinsic noise on proteins' distributions. From the deterministic system, one can see that the mean amount of target protein is proportional to the amount of its unrepressed mRNA, that is those molecules not bound to miRNAs and free for translation. In presence of extrinsic noise the target mRNA can show bimodal distributions even without clearly having a double steady state coming from the deterministic system. We here investigate if this is the case for the protein distribution. A key factor to keep into account is the time scale of protein synthesis and degradation. In general, if the protein dynamics are fast, the protein distribution follows closely the one of the mRNA (see Figures 5A1,5B1, 5C1). Conversely, slower protein dynamics tend to filter out the intrinsic fluctuations of the mRNA and lead to narrower distributions (see the supplementary material for a detailed discussion of the case without extrinsic noise). This is a single cell effect, that is, for a given rate of miRNA transcription $k_s$, the corresponding protein distribution gets more peaked as the protein dynamics get slower. Then, also the protein distribution subject to extrinsic noise tends to concentrate around its mean. This feature bears different consequences on the protein distribution shape according to the specific structure of the mRNA distribution. If the mRNA distribution is bimodal, slower proteins will have a distribution condensing around their mean which is located close to the unrepressed peak. They will therefore preferentially display the unrepressed phenotype and may completely lose their bimodal structure (see Figure 5A2). Bimodality can persist for strongly bimodal mRNA distribution because the noise reduction mechanism is acting at the single cell level so that it cannot overcome the effects of the extrinsic noise (see Figure 5B1,B2).

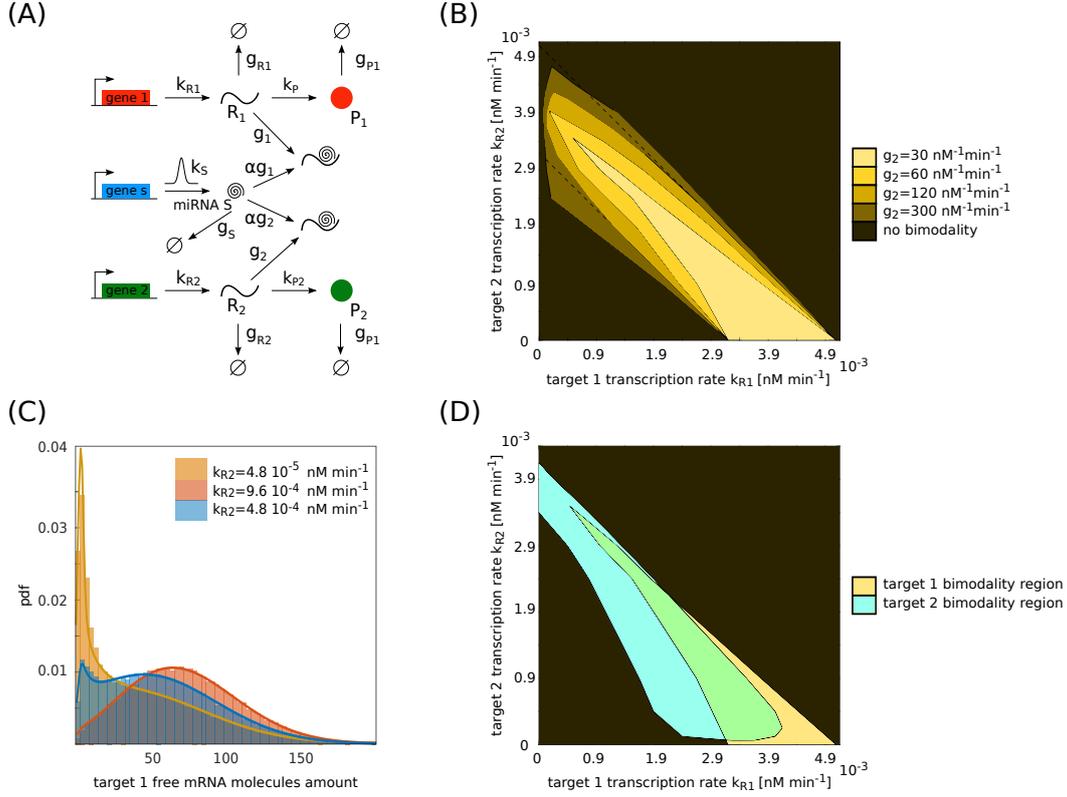

Figure 4: **Competition between two targets of the same miRNA.** (A) Reference circuit including extrinsic noise for the case of two genes competing for the same miRNA. $k_S$ is the miRNA transcription rate. $k_{R1}$ and $k_{R2}$ are the mRNA transcription rates and $g_{R1}$ and $g_{R2}$ are the mRNA degradation rates of target 1 and 2 respectively. $k_{P1}$ and $k_{P2}$ are the translation rates and $g_{P1}$ and $g_{P2}$ are the degradation rates of protein 1 and 2 respectively. $g1$ and $g2$ are the miRNA interaction rates with target 1 and 2. $\alpha$ is the fraction of miRNAs that are not recycled after binding to the mRNAs. (B) Phase diagram for the bimodality of the target $R_1$ for a fixed level of extrinsic noise ($\sigma = 4.8 \times 10^{-4}$ nM min$^{-1}$), small miRNA/target 1 interaction strength ($g_1 = 1.2 \times 10^2$ nM$^{-1}$ min$^{-1}$) and different miRNA/target 2 interaction strengths $g_2$. The other parameters are as follows: $k_{R1}$ and $k_{R2}$ range from 0 nM min$^{-1}$ to $5.1 \times 10^{-3}$ nM min$^{-1}$, $\overline{k}_S = 1.2 \times 10^{-3}$ nM min$^{-1}$, $g_S = 1.2 \times 10^{-2}$ min$^{-1}$, $g_{R1} = g_{R2} = 2.4 \times 10^{-2}$ min$^{-1}$, $k_{P1} = k_{P2} = 6.0$ min$^{-1}$, $g_{P1} = g_{P2} = 2.4 \times 10^{-2}$ min$^{-1}$, $\alpha = 0.5$. (C) Explanatory example of how it is possible to modulate target 1 distribution increasing the expression of target 2 for small interaction strength between miRNA and targets ($g_1 = 1.2 \times 10^2$ nM$^{-1}$ min$^{-1}$, $g_2 = 30$ nM$^{-1}$ min$^{-1}$). The extrinsic noise is here $\sigma = 2.4 \times 10^{-4}$ nM min$^{-1}$. The other parameters are as in (B). (D) Intersection between the bimodality phase diagrams of both targets for $g_1 = 1.2 \times 10^2$ nM$^{-1}$ min$^{-1}$ and $g_2 = 30$ nM$^{-1}$ min$^{-1}$. The other parameters are as in (B).

For a repressed (unimodal) mRNA distribution the mode is far from the mean, so that narrowing around the mean implies the rise of a second (unrepressed) peak. For moderately slower dynamics (see Figure 5C2) the protein distribution may be bimodal and for even slower ones it will be unimodal close to its mean (see Figure 5C3).

Altogether these results suggest that slow proteins promote the expression closer to the mean of the corresponding mRNA distribution. This may be sufficient to remove the bimodal feature of the protein distribution or not depending on the interplay among the amplitude of the extrinsic noise, the coupling between target and miRNA and the transcription rates.

## IV. DISCUSSION AND CONCLUSION

Previous studies pointed out the relevance of extrinsic noise in molecular networks in shaping cell decision making and differentiation [1, 2]. Gene expression levels can be normally observed into bimodal distributions, whose modes correspond to different physiological states of the system [3–8]. In this work we addressed the question on the role of extrinsic noise in shaping bimodal gene distributions in the context of miRNA-mediated regulation, both with stochastic modelling and simulations.

MiRNAs may induce bimodality in the expression of their target genes simply due to peculiar titrative interactions [11, 12]. In a system with pure intrinsic noise, such bimodal distributions can be observed in conditions of high miRNA-target interaction strength and for a small range of target transcription rates. Indeed, the binding and unbinding reactions between miRNA and target in condition of quasi equimolarity let the target jump from the bound to the unbound state, giving rise to bimodal distributions. This bimodality is observed at a single-cell level: every single cell can indeed switch from one state to the other so that at a given time part of the population will be "bound" and part "unbound".

We showed that introducing some extrinsic noise on the miRNA transcription widens the range of the target transcription rate for which one observes target bimodality.

In this case the bimodal distributions arise at the population level, made of several cells that are heterogeneous with respect to the miRNA expression and therefore amount. Hence the bimodality comes out from the superposition of those unimodal distributions describing the single cells, i.e., each one obtained for a different value of miRNA transcription rate.

Interestingly, a high miRNA-target interaction strength is not necessary to obtain a population-induced bimodal distribution. We showed that extrinsic noise and miRNA-target interaction strength act at similar levels with respect to the bimodality. The interaction strength between miRNA and target in our model takes into account possible different number of miRNA binding sites on the mRNA target sequence. Since the miRNA repression on a given target is usually small and possibly diluted over multiple targets, our results suggest that some extrinsic noise can compensate for a low interaction strength in order to obtain differentially expressed phenotypes.

Since every single miRNA may have many different targets that in turn compete for the shared pool of miRNAs, a change in the expression level of one of them may alter the expression of the others. If one of these targets is bimodally distributed, then the bimodality may be influenced by the expression level of the other miRNA competitors. We modelled this scenario considering two targets in competition for the same miRNA and showed that cross regulation is possible even in case of small miRNA-target interaction strength if some extrinsic noise is present. In particular, different targets may cross-regulate each other's bimodal distributions and their interplay is pivotal in stabilising the presence of single phenotypes. This suggests that even if the miRNA repression is low and diluted over a network of multiple targets, the noisy environment makes cross-regulation among them possible.

The aim of regulatory systems is to control protein expression. Concerning the effect of extrinsic noise on their distribution, we showed that depending on the time scales of protein synthesis and degradation, the protein distribution may suppress or amplify the bimodality at the level of mRNA.

Altogether our results suggest that the coupling between extrinsic noise and threshold behaviour represents an efficient mechanism to obtain bimodal phenotypes without the need of fine tuning the rates of reactions which was required for the case of intrinsic noise only.

However, questions on the biological relevance of extrinsic noise may arise, especially about the perks of this high variability of different cells within a same tissue. Through our analysis, we observed that in both one-target and multiple miRNA-targets circuits, even in presence of extrinsic noise the region of unimodality is still quite wide (see Fig. 3A, 4B and D): the system is still able to noise buffer the extrinsic fluctuations and channel one only final phenotype into the protein branch within a wide range of parameters. In the remaining part, where the extrinsic noise is not buffered, other downstream mechanisms here neglected could help taking cell fate decision and hence lead the cell in a repressed or unrepressed state. In this case, the bimodal distribution induced by extrinsic noise would work as "enhancer" of cell-cell variability.

A feasible strategy to test the experimental validity of these theoretical results involves building ad-hoc synthetic circuits made of miRNA targets tagged with fluorescent labels, as previously done in [12, 36], and performing transfection experiments. By analysing the fluorescence patterns of the targets interacting with the miRNA throughout the entire population of cells, the shape of the target distributions can then be extracted and compared to the theoretical predictions. MiRNAs are differentially expressed in different tissues and we expect also the amplitudes of their extrinsic fluctuations to vary consequently. Then, repeating the experiment in cells derived from different tissues would allow to study cells exposed to different levels of extrinsic noise. This could be used to contrast the observed dependence of the phenotypes on the level of extrinsic noise against the one predicted by the model. It would be then interesting to investigate how to artificially induce miRNAs fluctuations. This would provide a tool to control the arising of bimodality in targets distributions and to study its consequences in disease-related contexts

## V. FUNDING

This work was supported by the Francis Crick Institute which receives its core funding from Cancer Research UK (FC001317), the UK Medical Research Council (FC001317) and the Wellcome Trust (FC001317) to S. G.; the Swedish Science Council (Vetenskapsrådet) [Grant No. 621-2012-2982] to M.D.G. and C.B.

## VI. ACKNOWLEDGEMENTS

The authors would like to thank Chiara Enrico Bena and Raffaele Marino for useful discussions and Andrea De Martino, Matteo Marsili, Olivier Martin, Andrea Pagnani, Guillaume Salbreux, Philipp Thomas and Riccardo Zecchina for enlightening suggestions and critical reading of this manuscript. M.D.G. and C.B. wish to thank Nordita for the hospitality during part of this work.

### 1. Conflict of interest statement.

None declared.

---

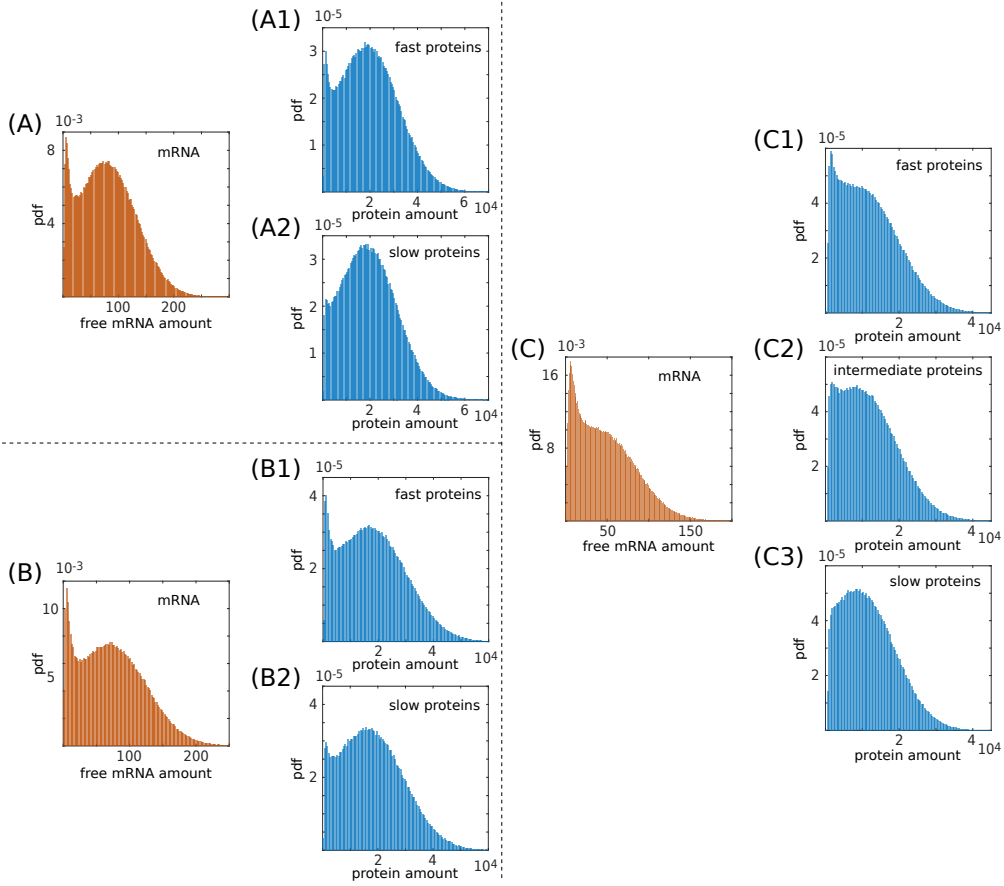

Figure 5: **Proteins half-life and bimodality.** In this panel three conditions (A), (B) and (C) in which the shape of the protein distribution is altered by an increased protein stability are reported. Histograms are the result of numerical simulations. In orange are represented the free mRNA distributions and in blue the protein distributions corresponding to different levels of scale separation between the mRNA and protein dynamics. Fast proteins distributions are obtained for a protein half life comparable to the mRNA one; in this condition the state of the protein is copying the mRNA one and the distributions almost coincide. Slow protein distributions are obtained for a protein half life up to 10 times longer than the mRNA one; as a consequence of the higher protein stability different outcomes can be achieved, depending on the level of extrinsic noise, the miRNA-target interaction strength and the closeness to the threshold ($k_R$). Starting with a well defined bimodal distribution (A1) and (B1), for a fixed level of extrinsic noise, the repressed peak can be buffered (A2) or not (B2) depending on the value of $k_R$. If the initial distribution is unimodal repressed (C1), for a given range of parameters, it can be converted into a unimodal unrepressed (C3), crossing a bimodal state (C2), by increasing the protein stability. In (A) the parameters are $k_R = 3.1 \times 10^{-3}$ nM min$^{-1}$, $\overline{k}_S = 1.2 \times 10^{-3}$ nM min$^{-1}$, $\sigma = 2.4 \times 10^{-4}$ nM min$^{-1}$, $g_S = 1.2 \times 10^{-2}$ min$^{-1}$, $g_R = 2.4 \times 10^{-2}$ min$^{-1}$, $g = 1.2 \times 10^2$ nM$^{-1}$ min$^{-1}$, $\alpha = 0.5$, $k_P = 6.0$ min$^{-1}$, $g_P = 2.4 \times 10^{-2}$ min$^{-1}$ for (A1) and $k_P = 6.0 \times 10^{-1}$ min$^{-1}$, $g_P = 2.4 \times 10^{-3}$ min$^{-1}$ for (A2). In (B) the parameters are $k_R = 3.0 \times 10^{-3}$ nM min$^{-1}$, $\overline{k}_S = 1.2 \times 10^{-3}$ nM min$^{-1}$, $\sigma = 2.4 \times 10^{-4}$ nM min$^{-1}$, $g_S = 1.2 \times 10^{-2}$ min$^{-1}$, $g_R = 2.4 \times 10^{-2}$ min$^{-1}$, $g = 1.2 \times 10^2$ nM$^{-1}$ min$^{-1}$, $\alpha = 0.5$, $k_P = 6.0$ min$^{-1}$, $g_P = 2.4 \times 10^{-2}$ min$^{-1}$ for (B1) and $k_P = 6.0 \times 10^{-1}$ min$^{-1}$, $g_P = 2.4 \times 10^{-3}$ min$^{-1}$ for (B2). In (C) the parameters are $k_R = 3.1 \times 10^{-3}$ nM min$^{-1}$, $\overline{k}_S = 1.4 \times 10^{-3}$ nM min$^{-1}$, $\sigma = 1.7 \times 10^{-4}$ nM min$^{-1}$, $g_S = 1.2 \times 10^{-2}$ min$^{-1}$, $g_R = 2.4 \times 10^{-2}$ min$^{-1}$, $g = 1.2 \times 10^2$ nM$^{-1}$ min$^{-1}$, $\alpha = 0.5$, $k_P = 6.0$ min$^{-1}$, $g_P = 2.4 \times 10^{-2}$ min$^{-1}$ for (C1), $k_P = 3.0$ min$^{-1}$, $g_P = 1.2 \times 10^{-2}$ min$^{-1}$ for (C2) and $k_P = 1.2$ min$^{-1}$, $g_P = 4.8 \times 10^{-3}$ min$^{-1}$ for (C3). The ratio between $k_P$ and $g_P$ is always kept constant in order to maintain fixed the mean of the protein distributions. Note that we do not present the analytic curves for such cases as the approximation fails at capturing subtle features as the (dis-)appearance of a small peak (see SI).

# On the role of extrinsic noise in microRNA-mediated bimodal gene expression - Supplementary Information

Marco Del Giudice, Stefano Bo, Silvia Grigolon, Carla Bosia

## I.  COMPARISON BETWEEN DIFFERENT ANALYTICAL APPROXIMATIONS

In the Main Text we showed how to obtain the steady state distribution $P(n_S, n_R, n_P|k_S)$ conditioned on a specific miRNA transcription rate by a system-size expansion [1]. Alternatively, one could resort to a gaussian approximation of the moments as in [2] which consists in closing the hierarchy of the equations for the moments by expressing the third one as a combination of means and covariances. The simplest way to actually produce the shape of the distribution $P(n_S, n_R, n_P|k_S)$ is to make the additional assumption that it is gaussian with averages and covariances given by the moment closure scheme. Both approximations can be convoluted with the distribution of $k_S$ to produce the complete distribution $P(n_S, n_R, n_P)$.

In this section we compare the results of the two approximations. For the plots in Figure 2 and 4 in the Main Text the two approximations give rise to basically indistinguishable curves (not shown) in agreement with the simulation results. However, they may fail to predict subtle details in the bimodal distribution, such as the initial emergence of the unrepressed peak addressed in Fig. 5 in the Main Text. For such parameters, the two approximations differ more markedly (see Fig. S1). In general, the gaussian approximation is more accurate in predicting the shape of the protein distribution in case of high protein stability. This observation can be traced back to the fact that the gaussian approximation over-performs the van Kampen one in describing the mean value of mRNA and proteins (see Fig. S1 D). As discussed in the Main Text, proteins with slower dynamics narrow their distribution around their mean so that predicting the mean accurately becomes more relevant. In Figure S1 we show a comparison between the two approximations in relation to the histograms represented in Fig. 5 in the main text.

However, when the timescales of mRNAs and proteins are comparable there seems to be no general rules on when one approximation is to be preferred. For the investigation of subtle features of the distributions, the approximation should be just used as a qualitative reference to be validated by simulations.

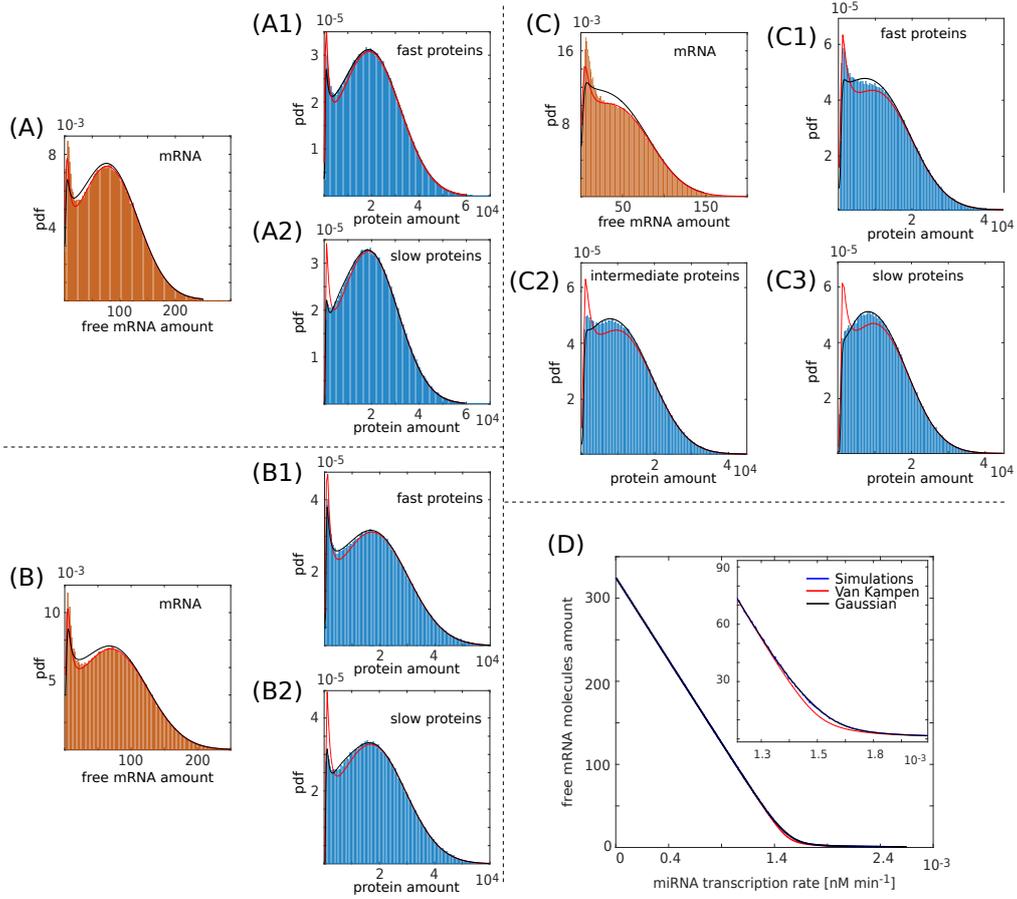

**FIG. S1: Comparison between Van Kampen and gaussian approximations**. In (A-C) mRNA and protein distributions for unstable and stable proteins are shown together with the two approximations. In (D) the mean number of mRNA molecules as a function of the miRNA transcription rate is shown, the blue line corresponds to numerical simulations, while the red and black one are the theoretical predictions obtained through the Van Kampen and Gaussian approximation respectively. The inset is a zoom of the region where the difference between the two approximations is more evident. In (A) the parameters are $k_R = 3.1 \times 10^{-3}$ nM min$^{-1}$, $\bar{k}_S = 1.2 \times 10^{-3}$ nM min$^{-1}$, $\sigma_S^2 = 2.4 \times 10^{-4}$ nM min$^{-1}$, $g_S = 1.2 \times 10^{-2}$ min$^{-1}$, $g_R = 2.4 \times 10^{-2}$ min$^{-1}$, $g = 1.2 \times 10^2$ nM$^{-1}$ min$^{-1}$, $\alpha = 0.5$, $k_P = 6.0$ min$^{-1}$, $g_P = 2.4 \times 10^{-2}$ min$^{-1}$ for (A1) and $k_P = 6.0 \times 10^{-1}$ min$^{-1}$, $g_P = 2.4 \times 10^{-3}$ min$^{-1}$ for (A2). In (B) the parameters are $k_R = 3.0 \times 10^{-3}$ nM min$^{-1}$, $\bar{k}_S = 1.2 \times 10^{-3}$ nM min$^{-1}$, $\sigma_S^2 = 2.4 \times 10^{-4}$ nM min$^{-1}$, $g_S = 1.2 \times 10^{-2}$ min$^{-1}$, $g_R = 2.4 \times 10^{-2}$ min$^{-1}$, $g = 1.2 \times 10^2$ nM$^{-1}$ min$^{-1}$, $\alpha = 0.5$, $k_P = 6.0$ min$^{-1}$, $g_P = 2.4 \times 10^{-2}$ min$^{-1}$ for (B1) and $k_P = 6.0 \times 10^{-1}$ min$^{-1}$, $g_P = 2.4 \times 10^{-3}$ min$^{-1}$ for (B2). In (C) the parameters are $k_R = 3.1 \times 10^{-3}$ nM min$^{-1}$, $\bar{k}_S = 1.4 \times 10^{-3}$ nM min$^{-1}$, $\sigma_S^2 = 1.7 \times 10^{-4}$ nM min$^{-1}$, $g_S = 1.2 \times 10^{-2}$ min$^{-1}$, $g_R = 2.4 \times 10^{-2}$ min$^{-1}$, $g = 1.2 \times 10^2$ nM$^{-1}$ min$^{-1}$, $\alpha = 0.5$, $k_P = 6.0$ min$^{-1}$, $g_P = 2.4 \times 10^{-2}$ min$^{-1}$ for (C1), $k_P = 3.0$ min$^{-1}$, $g_P = 1.2 \times 10^{-2}$ min$^{-1}$ for (C2) and $k_P = 1.2$ min$^{-1}$, $g_P = 4.8 \times 10^{-3}$ min$^{-1}$ for (C3). In (D) the parameters are $k_R = 3.1 \times 10^{-3}$ nM min$^{-1}$, $g_S = 1.2 \times 10^{-2}$ min$^{-1}$, $g_R = 2.4 \times 10^{-2}$ min$^{-1}$, $g = 1.2 \times 10^2$ nM$^{-1}$ min$^{-1}$, $\alpha = 0.5$, $k_S$ ranges from 0 to $2.6 \times 10^{-3}$ nM min$^{-1}$.

## II. THE COEFFICIENT OF VARIATION AS AN INDICATOR OF BIMODALITY LOSS IN CASE OF PURE INTRINSIC NOISE

The comparison of coefficient of variations ($CV$) are common measures of noise buffering in molecular networks. Here we focus on the comparison between the coefficient of variations of the targets ($CV_R$) and the proteins ($CV_P$) in case of pure intrinsic noise. If there is noise buffering, i.e. the amount of noise is reduced by the production/degradation of proteins, the protein CV will be smaller than the target one. Conversely, if there is no noise buffering, the two CVs are exactly the same. To give a prediction on the values of the parameters for which noise buffering is present, one can use van Kampen's system-size expansion's solution and build up analytical expressions for the coefficient of variations of targets and proteins. Even by fixing all parameters but miRNA transcription rates and proteins transcription and degradation rates, the expression for the coefficient of variations are quite cumbersome. To understand qualitatevely the behaviour of the Coefficient of Variation as a function of the degradation rate of the proteins, one can fix all the other parameters and compute the expression of the CV. Imposing $CV_P - CV_R \leq 0$ allows to quantify the range of values of $g_P$ to have noise buffering in the final protein channel. In particular we found that $g_P \leq 0.9$ min$^{-1}$ for the parameters' choice as in Fig. 5A-A1, i.e., for values of $g_P$ smaller than a given threshold and hence for long-living proteins.

---

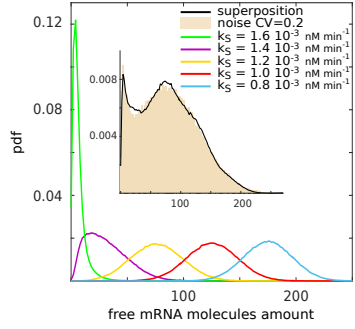

**FIG. S2: Comparison between the bimodal mRNA noisy distribution and the weighted superposition of distributions obtained without noise for different miRNA transcription rates.** The parameters are the following: $k_R = 3.1 \times 10^{-3}$ nM min$^{-1}$, $g_S = 1.2 \times 10^{-2}$ min$^{-1}$, $g_R = 2.4 \times 10^{-2}$ min$^{-1}$, $g = 1.2 \times 10^2$ nM$^{-1}$ min$^{-1}$, $k_P = 6.0$ min$^{-1}$, $g_P = 1.2 \times 10^{-2}$ min$^{-1}$ and $\alpha = 0.5$. In the main plot the different mRNA distributions correspond, from left to right, to $k_S = 1.7 \times 10^{-3}, 1.4 \times 10^{-3}, 1.2 \times 10^{-3}, 9.5 \times 10^{-4}, 7.1 \times 10^{-4}$ nM min$^{-1}$. In the inset, the mRNA histogram is the result of miRNA transcription rates picked from a gaussian distribution with mean $\bar{k}_S = 1.2 \times 10^{-3}$ nM min$^{-1}$ and standard deviation $\sigma = 2.4 \times 10^{-4}$ nM min$^{-1}$. The black line is the result of the weighted superposition of the distributions represented in the main plot.

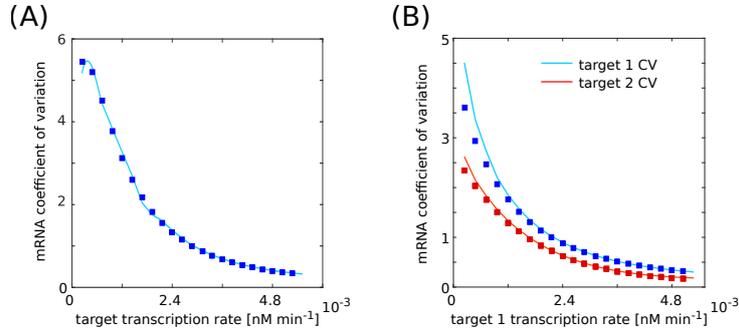

**FIG. S3: Analytical prediction for the coefficient of variation.** Analytical predictions for the target coefficient of variation in case of one (A) or two (B) targets. In (A) the parameters are $\bar{k}_S = 1.2 \times 10^{-3}$ nM min$^{-1}$, $\sigma = 4.8 \times 10^{-4}$ nM min$^{-1}$, $g_S = 1.2 \times 10^{-2}$ min$^{-1}$, $g_R = 2.4 \times 10^{-2}$ min$^{-1}$, $g = 1.2 \times 10^2$ nM$^{-1}$ min$^{-1}$, $k_P = 6.0$ min$^{-1}$, $g_P = 1.2 \times 10^{-2}$ min$^{-1}$, $\alpha = 0.5$. $k_R$ ranges from $2.4 \times 10^{-4}$ nM min$^{-1}$ to $5.2 \times 10^{-3}$ nM min$^{-1}$. In (B) the parameters are $\bar{k}_S = 1.2 \times 10^{-3}$ nM min$^{-1}$, $\sigma = 4.8 \times 10^{-4}$ nM min$^{-1}$, $g_S = 1.2 \times 10^{-2}$ min$^{-1}$, $g_{R1} = g_{R2} = 2.4 \times 10^{-2}$ min$^{-1}$, $g_1 = 1.2 \times 10^2$ nM$^{-1}$ min$^{-1}$, $g_2 = 30$ nM$^{-1}$ min$^{-1}$, $k_{P1} = k_{P2} = 6.0$ min$^{-1}$, $g_{P1} = g_{P2} = 1.2 \times 10^{-2}$ min$^{-1}$, $\alpha = 0.5$, $k_{R2} = 9.5 \times 10^{-4}$ nM min$^{-1}$. $k_{R1}$ ranges from $2.4 \times 10^{-4}$ nM min$^{-1}$ to $5.2 \times 10^{-3}$ nM min$^{-1}$.

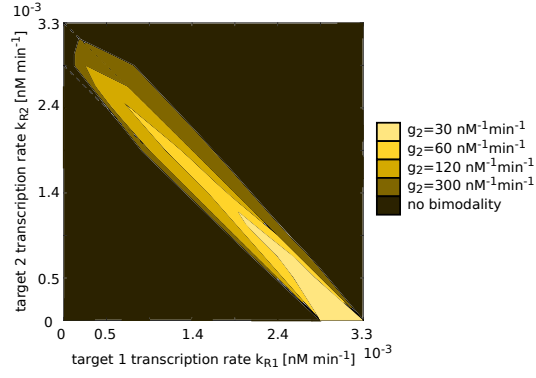

**FIG. S4: Bimodality phase diagram.** The plot shows the bimodality phase diagram for the mRNA 1 in a system with two targets competing for the same miRNA. The parameters here used are the following: $\overline{k}_S = 1.2 \times 10^{-3}$ nM min$^{-1}$, $\sigma_S = 2.4 \times 10^{-4}$ nM min$^{-1}$, $g_1 = 1.2 \times 10^2$ nM$^{-1}$ min$^{-1}$, $k_{R1}$ and $k_{R2}$ range from 0 nM min$^{-1}$ to $5.1 \times 10^{-3}$ nM min$^{-1}$, $g_S = 1.2 \times 10^{-2}$ min$^{-1}$, $g_{R1} = g_{R2} = 2.4 \times 10^{-2}$ min$^{-1}$, $k_{P1} = k_{P2} = 6.0$ min$^{-1}$, $g_{P1} = g_{P2} = 2.4 \times 10^{-2}$ min$^{-1}$, $\alpha = 0.5$.